# A Deep, Forgetful Novelty-Seeking Movie Recommender Model

Ruomu Zou
Phillips Exeter Academy
Exeter, New Hampshire 03833
Email: rzou@exeter.edu

## Abstract

As more and more people shift their movie watching online, competition between movie viewing websites are getting more and more intense. Therefore, it has become incredibly important to accurately predict a given user's watching list to maximize the chances of keeping the user on the platform. Recent studies have suggested that the novelty-seeking propensity of users can impact their viewing behavior. In this paper, we aim to accurately model and describe this novelty-seeking trait across many users and timestamps driven by data, taking into consideration user forgetfulness. Compared to previous studies, we propose a more robust measure for novelty. Our model, termed Deep Forgetful Novelty-Seeking Model (DFNSM), leverages demographic information about users, genre information about movies, and novelty-seeking traits to predict the most likely next actions of a user. To evaluate the performance of our model, we conducted extensive experiments on a large movie rating dataset. The results reveal that DFNSM is very effective for movie recommendation.

## Key Words

# Table of Contents

# 1 Introduction

With the growth in popularity of video sharing websites, the role online movie watching plays in our lives has grown steadily. It was reported that many users have transitioned from using traditional platforms like television to the internet for their movie-viewing [5]. This phenomenon has created a wealth of consumer data on what movies a user watches, and have allowed for much more accurate, personalized, and time-sensitive recommendations for all viewers. Recommending movies that users are more likely to watch are important as it keeps them on the platform for more sessions.

Fundamentally, this problem is about recommending the most likely movies a user might watch in the next timestep, which can be generalized as the most probable actions a user might take. Many classical methods only considering users and items, such as collaborative filtering (CF) and content-based filtering, compute their similarity to determine the recommendations [8,9]. More recently, deep learning has also been integrated into the field of recommendation. These include techniques that aid CF such as Neural Collaborative Filtering, and Neural Network Matrix Factorization, both of which adds more complexity and non-linearity to traditional CF [10, 11]. These techniques, while incredibly powerful, ignore the impact of time and user novelty-seeking. In other words, they do not consider the sequence in which users choose the items, and thus assumes that every user always wants to be recommended the same type of item.

To solve this problem, many new methods operating on user action sequences, or sequential representations of choices made by a specific user, emerged. Many of these methods try to quantify the idea of creating a more diverse set of recommendations, such as Sequential Hierarchical attention Network (SHAN), which creates separate representations for the short-term and long-term interests of a user and makes predictions accordingly [12]. However, among all these quantifications, perhaps the most all-encompassing is measuring user novelty-seeking at different times [2]. Novelty-seeking is a phenomenon where "through some internal drive or motivating force the individual is activated to seek out novel information" [1]. Studies suggest that this trait is genetic and might have evolutionarily encouraged exploration in humans and thus allowed us to form a more thorough understanding of the world [1]. Novelty-seeking is also regarded as a basic requirement for humans [7].

In the context of this paper, we consider the novelty-seeking from two perspectives. The first in relation to each action the user can possibly take and indicates how different this action is from previous actions taken by the user at one timestamp. In the case of movie recommendation action novelty is computed in relation to the tags of each film. The second is derived from action novelty and represents how novelty-seeking a user might be at that particular time.

There have been many previous studies aimed at quantifying the effect of novelty-seeking on consumer behavior [2,3]. For example, Zhang et al. proposed a framework named Novelty-Seeking Model (NSM). It uses a dynamic choice novelty (DCN) matrix to store the novelty rankings of actions, and Gibbs sampling for prediction. However, their representation of action novelty only stores integers that indicate relative ranking, or how new one action is compared to others. This neglects the true nuance of novelty, as it is oblivious to how much newer some actions are compared to others. For example, action A might be ten times more novel than action B, but only two times newer than action C, and NSM can only output a ranking of these actions, effectively treating the changes in novelty from one action in the ranking to another as a constant value. NSM also computes a single DCN matrix for each user, which takes into account every single action this user made since registration. In practice, this approach is not only inefficient, but also inaccurate. Most users will not necessarily remember very action they took years ago, and a long-ago action will not make a similar action in the present too much less novel.

We propose the Deep Forgetful Novelty-Seeking Model (DFNSM). DFNSM is based on the following observations about how novelty-seeking relates to movies watched. Firstly, the choice made at time T is most significantly impacted by action novelty values calculated from choices made between times T-k and T-1, inclusive, where k is an integer. In other words, there is no need to consider the impact of actions a long time ago when calculating action novelty, because the user has likely already forgotten many aspects of it and experiencing a similar action in the present will still be very novel. Therefore, we do not need to store all actions since the initial one, instead only k actions are needed (hence the word "forgetful"), massively increasing the efficiency of the model and at the same time eliminating potentially misleading information. Secondly, how much more novel an action is from another is important and measuring the novelty of actions simply as a ranking neglects this.

DFNSM combines deep learning models like Multi-layer Perceptrons (MLP) and Convolutional Neural Networks (CNN) with the concept of using novelty-seeking values to aid the prediction of the next movie viewed by a user. We leverage information on user demographics (age, job, etc.), movie content (title, tags, etc.), and tag-specific user novelty-seeking values to predict the rating a user might give any movie. A representation of the user is constructed through demographics, but not the past actions because demographics are less subject to change than a person's interests (as represented by previous actions). In summary, our contributions include:

- We combine different types of novelty-seeking traits to predict possibilities for the next action made by the user.
- We generalized the computation of action novelty by focusing on only k previous actions of a user for better accuracy and performance.
- We created a more robust measure of novelty by treating it as a scale, with each action ranging from most familiar to newest, instead of treating it merely as relative ranking and assuming that one action is always the same amount newer than the previous one.
- We analyzed the time impact of novelty seeking for different users in the domain of movie-viewing by choosing different k values.

## 2 Related Works

In this section we outline some publications that are related to the idea of novelty-seeking, various deep learning models, and the field of recommendation.

## 2.1 Novelty-Seeking Behavior

Novelty-seeking is an inherent trait to the human species, but it is a broad concept and many studies have been done to better describe it. Hirschman studied the effects of this trait on consumer behavior, and found that the seek for new information is one of the driving factors behind consumption of "information-rich" materials like magazines [1]. Her results implied that novelty-seeking is worth describing and quantifying to improve recommender systems.

## 2.2 Deep-Learning Models

With the rapid increase in computation power, deep learning algorithms are becoming more and more popular. Among them are two powerful kinds of neural networks: MLPs and CNNs [4,6]. It has been shown that all deep neural networks are universal function approximators, meaning that an arbitrarily complex network can approximate any function will an arbitrarily small amount of error [15]. An MLP is a network consisting of many layers of fully connected nodes, with each node taking in as input the outputs of all nodes in the previous layer and computing a weighted sum on them before adding a bias value. That value is then passed through some non-linear function [4]. A CNN can be seen as a special case of an MLP, because it is specifically used on data where the most relevant semantic information is contained in the relationships between neighboring input cells, such as images or text, while MLPs simply try to capture any relationship between any number of input cells [6].

## 2.3 Recommender Systems

To explore the recommendation problem, many models have been proposed. Several of the most iconic and effective ones are the Collaborative Filtering (CF), and User-based Filtering algorithms. CF is a set of techniques which predict user ratings/interest (more generally, affinities) with certain items based on a) this user's interaction with other items, and b) this item's affinities with other similar users. The idea behind CF is that the interests between users who rate identical items similarly are also similar [8]. Content-based filtering, on the other hand, creates a representation of user interests based on the items rated in the past, using information such as item tags etc. for prediction [9]. Deep learning has also been heavily integrated into these traditional techniques. For example, He et al. proposed the Neural collaborative filtering model, and Dziugaite et al. proposed the Neural network matrix factorization model. Both of these add MLP networks to various parts of CF. The added deep neural networks allow for significantly more complex predictive capacity and allows the model to accurately approximate non-linear relations from the non-linear activation layers of the MLPs [10,11].

Models operating on user action sequences have also emerged. Ying et al. proposed a model named Sequential Recommender System based on Hierarchical Attention Network (SHAN). SHAN aims to find both the long-term and short-term interests for a given user to aid in recommendation.

There have also been many models combining the ideas of novelty-seeking and recommendation. Zhang et al. proposed the NSM framework. NSM frames action novelty as a matrix of values indicating how novel each action is at each time and uses Gibbs sampling for inference [2]. A subsequent study done by Zhang et al. on restaurant recommendation uses a conditional random field for predicting a Boolean value indicating if a user would like to visit a previously unfamiliar restaurant [3].

# 3 Deep Forgetful Novelty-Seeking Model (DFNSM)

In this section, we first formulate the next-action recommendation problem explored in this paper. We then introduce our measure of action and user novelty before presenting the Deep Forgetful Novelty-Seeking Model (DFNSM).

## 3.1 Problem Formulation

Let $U$ be a set of users, $M$ be a set of movies each user can chose from, and $I$ be a set of tags describing the movies (such as "comedy", "fantasy", etc.). $I_m$ is then a subset of $I$ containing the individual tags for an action $m \in M$. For each user $u \in U$ let $S^u = \{A_1^u, A_2^u, A_3^u, \ldots, A_T^u\}$, where T is the number of total actions made by user u, and $A_{1\ldots T}^u$ is a set of tuples representing the viewing actions made. $A_{t \in T}^u$ consists of the title of movie $m \in M$ chosen, the set $I_m$, action novelty values for each tag in $I_m$ at time $t$, and the rating user u gave the movie.

We aim to predict this rating through the other elements of the action. In other words, given the ANI values of a user, this user's demographic information, and information about every movie the user can chose, we predict ratings this user is likely to give to these movies and produce the final recommendations by sorting these ratings.

## 3.2 Quantifying Novelty-Seeking by Formulating the Action Novelty Index (ANI) Matrix and the User Novelty-Seeking Index (UNI)

In this subsection we introduce our definitions for the ANI matrix and the UNI. We start by defining the ANI matrix and then use it to formulate UNI.

### 3.2.1 Formulating the ANI Matrix

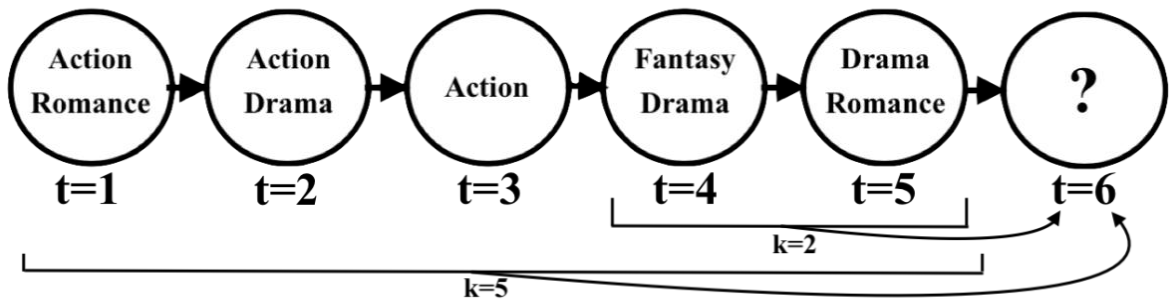

Figure 1: A sample user action sequence, with each action represented by tags of the movie viewed. How k values act in computing ANI matrices are also shown

| | Action | Romance | Drama | Fantasy |
|---|---|---|---|---|
| 1 | 1 | 1 | 1 | 1 |
| 2 | 0.5 | 0.5 | 1 | 1 |
| 3 | 0.33 | 0.5 | 0.5 | 1 |
| 4 | 0.33 | 1 | 0.5 | 1 |
| 5 | 0.5 | 1 | 0.5 | 0.5 |
| 6 | 1 | 0.5 | 0.33 | 0.5 |

k=2

| | Action | Romance | Drama | Fantasy |
|---|---|---|---|---|
| 1 | 1 | 1 | 1 | 1 |
| 2 | 0.5 | 0.5 | 1 | 1 |
| 3 | 0.33 | 0.5 | 0.5 | 1 |
| 4 | 0.25 | 0.5 | 0.5 | 1 |
| 5 | 0.25 | 0.5 | 0.33 | 0.5 |
| 6 | 0.25 | 0.33 | 0.25 | 0.5 |

k=5

Figure 2: ANI matrices with k values as two and five computed from sequence in Figure 1.

Unlike previous studies, our model does not consider the entire action sequence of a user when computing action novelty, because, due to user forgetfulness, long-ago actions do not strongly impact what is novel at the present moment. In fact, they might even add misleading information, as will be elaborated on later in this subsection. We only consider the k most recent prior actions from the current one when computing action novelty, which we will define as how novel each possible action is to the user at the present moment. In the case of movie recommendation when repeated viewing is rare, we compute novelty values for the tags of a given movie. Action novelty is stored in a matrix named the Action Novelty Index (ANI) matrix.

The ANI matrix for a specific user is a matrix of size $T \times |I|$ (length of action sequence times number of tags total). $ANI_{t,i}$ represents how novel tag $i \in I$ is at time $t \in T$. Figures 1 and 2 provides an example action sequence as well as two ANI matrices computed for different k values. As seen in Figure 1, at each timestamp the user is faced with $|M|$ many actions (each consisting of a set of tags $I_{m \in M}$), and choosing any one will indicate that the tags associated with it are slightly less novel to the user for k future timesteps. Formally, the definition of $ANI_{t,i}$ is given as

$$ANI_{t,i} = \frac{1}{\#x^i_{t-k,t-1} + 1}$$

where $\#x^i_{t-k,t-1}$ denotes the number of times tag i had appeared in actions numbering from t-k to t-1. When t-k is below zero, which happens when the user's action amount is below k, we simply compute ANI values from actions numbering zero to t-1.

Figures 1 and 2 also provide a visual example, albeit an exaggerated one, for the advantages of only using previous k actions of ANI computation. The user depicted clearly has an interest in action movies in the beginning, but after a number of actions (in this case, three) started to give up on the action genre and instead developed a passion for drama. Even though a user is hardly going to forget about an entire genre over the course of five actions (it is only an example), in real life these shifts could occur gradually and over much more actions and a much longer time, by the end of which many users will still find some of their long-ago favorite genres to be relatively novel. In the case of a smaller k value relative to the action sequence length, at time six the ANI matrix is already treating the action genre as completely novel, while a bigger k value relative to action sequence length still remembers the user's past passion for action. If the user has an exceedingly good memory of long-gone interests, the bigger k-value is more suitable, as it does not neglect their influences on the ANI. However, a fickler user demands a smaller k, as too much redundant information about past interests will be misleading and inefficient. Most of us lie in between these two extremes, and assuming that everyone has perfect memory of past actions, as in models like NSM, is an oversimplification [2]. The constant k which produces the highest accuracy for a user can therefore be seen as a measure of fickleness of interest and defining such a number gives our model unprecedented flexibility.

In summary, there are several major differences between the ANI matrix and the DCN matrix proposed by Zhang et al. [2]. Firstly, ANI only considers the k previous actions, instead of the entirety of an increasing action sequence. The DCN matrix can be seen as a special case of ANI, when the k value is equal to the total action sequence length of a user. Secondly, ANI values are a truer representation of novelty, as they operate not as a ranking, but as a numerical value for a scale of how new every action is. Notice that ANI values will not keep tending to zero as the user takes more actions because the amount of actions used to compute $\#x_{t-k,t-1}^i$ is constant for all $t > k$. Thirdly, ANI does not take into account transition frequencies between actions, mainly because in the domain of movies, the amount of transitions between tags tend to be high and even mask the effect of tag appearance frequency.

### 3.2.2 Formulating the UNI

With a definition for measuring action novelty at each timestamp, we can now derive the user novelty-seeking index (UNI). The UNI is a representation of how novelty-seeking a given user is at a specific timestamp. The higher this value is, the more likely this user is going to choose new actions. This value is entirely for interpretation purposes and does not add extra information for prediction, as it is derived entirely from the row of the ANI matrix at that time. Given $ANI_{t \in T, 1...|I|}$, the formula for computing the UNI at time t ($UNI_t$) is given as

$$UNI_t = \frac{1}{F(Norm(ANI_{t \in T, 1...|I|})) + 1}$$

where the function Norm of an array x is given as the Frobenius Norm and the function F of an array x is given as

$$F(x) = \sum_{c=0}^{|x|} x_c * log_2(x_c)$$

or a measure for the entropy [13,14]. The purpose of using entropy here is as a measurement for the "smoothness" of array. This smoothness is highly connected to the UNI because the more evenly distributed a user's action novelties are, the more novelty-seeking this user is at the moment. Likewise, if the user only focuses on one or several genres, values in $ANI_{t \in T, 1...|I|}$ are going to be relatively uneven. The entropy is chosen over the variance because the variance is based on a mean and does not handle situations where the user focuses on more than one genre.

## 3.3 Model

In this subsection we propose the architecture of our model, as well as crucial information about its inference procedure.

### 3.3.1 Model Architecture

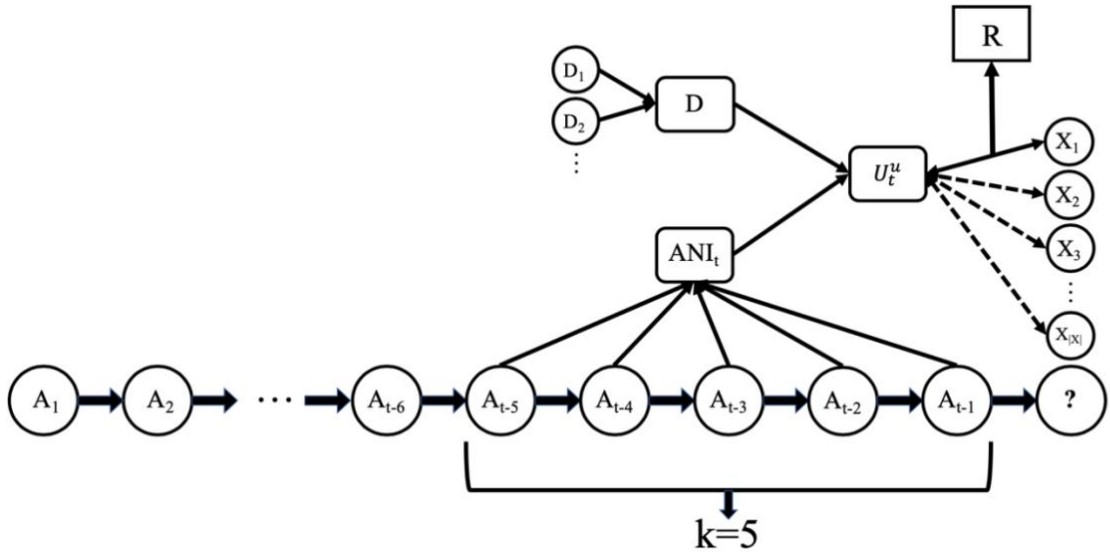

Figure 3: A graphical representation of the Deep Forgetful Novelty-Seeking Model

| Table 1: Summary tables of symbols in Figure 3 | |
|---|---|
| $A_t$ | Viewing action made at time t, for this model only the vector of tag information ($I_{m \in M}$) is passed |
| $ANI_t$ | A row of the ANI matrix indicating action novelty for each tag at time t |
| k | Scalar indicating how many past actions to use in computing $ANI_t$ |
| D | A vector containing demographic information on the user (age, job, gender etc.) |
| $U_t^u$ | A latent vector representing combined information about user u at time t |
| $X_m$ | A latent vector the same size as U representing information about movie $m \in M$ |
| R | Predicted rating the user will give a movie |

The basic idea behind DFNSM is to project information about the user at a timestamp and each movie onto the same latent dimensions ($U_t^u$ and $X_m$), and then computing a dot product to predict the rating this user is likely to give to a given movie. The movies can then be ranked according to this rating and the top several are recommended. $U_t^u$ is unique for each different user at each different timestep (due to different action novelty values), while $X_m$ only varies with different movies but is constant over time. For each movie, $X_m$ is derived from the movie title and tags as shown in Figure 4. The title provides unique information about content, while the tags offer genre information. Combined, they paint a clear and specific picture of each movie. We first run a convolution operation over the movie's title to extract textual information and embed its tags (summing the embeddings for each tag) [6]. We then run a fully connected MLP layer on these vectors to produce the final latent representation of the movie [4]. The fully connected layer gives complexity and non-linearity to the model, and therefore allows it to make better predictions.

Similarly, $U_t^u$ is derived from the demographic information (age, job, gender, etc.) of a user u combined with $ANI_t$ of this user. The combination is also done by means of a fully-connected MLP layer to increase predictive capacity [4]. Note that $U_t^u$ contains temporal information as well, due to the influence of $ANI_t$. $X_m$ and $U_t^u$ have the same number of dimensions, and the predicted rating is defined as their dot product.

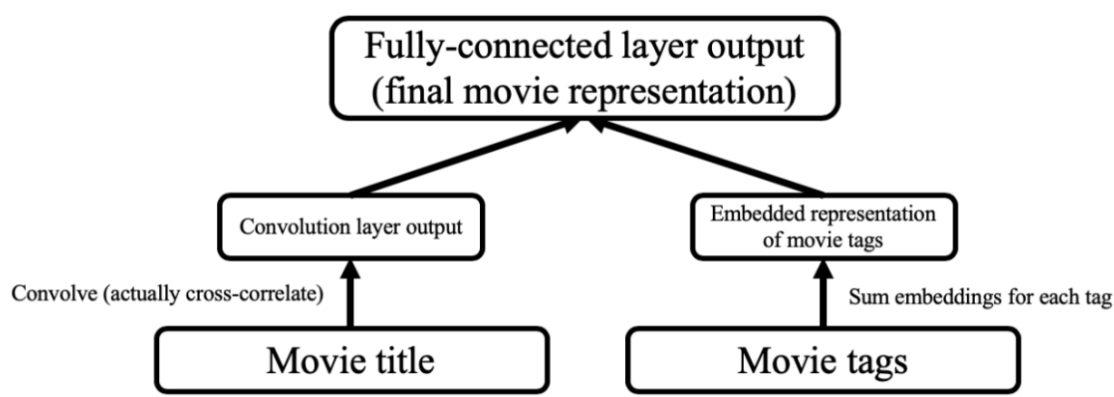

Figure 4: How $X_m$ values are computed

### 3.3.2 Inference

Algorithm 1: Inference algorithm for **Deep Forgetful Novelty-Seeking Model**

**Inputs:** Movie information (title, tags), user demographics, ANI matrices for users

**Output:** Model parameters Θ.

1. Draw Θ from a normal distribution with mean 0 and standard deviation 0.1.

2. **Repeat until** convergence

    a. Compute MSE loss for each batch of training examples

DFNSM trains via gradient descent, a popular training algorithm for neural networks. The target for optimization is a mean-squared-error (MSE) loss function given by

$$MSE(\hat{y}) = \frac{1}{n}\sum_{i=1}^{n} y_i - \hat{y}_i$$

where $\hat{y}$ is the model output and n denotes the number of training examples passed into the model (in training this value will be the batch size). Algorithm 1 gives the detailed procedure for inference.

# 4 Experiments

In this section, we first analyze our movie dataset. Based on this data, we analyze the performance of our model compared to those proposed in previous works. We also conduct an experiment to examine how different k-values impact the nDCG metric score for our model.

## 4.1 Dataset Analysis

| Table 2: Basic Statistics for the ml-1m Dataset | |
|---|---|
| User Number | 6,040 |
| Movies Number | 3,883 |
| Ratings Number | 1,000,209 |
| Mean User Action Sequence Length | 165.57 |
| Median User Action Sequence Length | 96 |
| Mean Tag Number/Movie | 1.65 |
| Male to Female Ratio | 2.53 |

We ran our model on the MovieLens 1M dataset (ml-1m) from GroupLens [16]. MovieLens is a non-commercial online movie recommender that recommends to users movies that they might like. The ml-1m dataset contains more than a million rating data points with timestamps from 6,040 users on around 3,883 movies [16]. Moreover, it contains basic demographics information about every user and basic genre information about movies. These characteristics make it an ideal dataset to be used with models like the DFNSM that are both context-aware (user and movie wise) and utilizes time sequences.

The original ml-1m dataset contains three subsets, respectively about user data, movie data, and rating data. The user subset is comprised of the following features: gender (either male or female for simplicity), age (in seven categories ranging from under 18 to above 56), and occupation (in 21 categories). In contrast, the movie subset only contains movie title and movie genre information in the form of tags. The ratings subsets connect the two other subsets by providing information about how users rated movies and at what times. To process the raw data into usable action sequences, we first merge all three subsets, dropping some undesired values, and then tokenize the movie tag and title information numerically. Most of our experiments are done on the first three users in the dataset, and our model is trained on the first twenty to save time.

The ratings information provides some interesting insights into consumer behavior and habits. Figure 5 and 6 illustrates the frequencies of the ratings most users tend to give. Overall, there is a significant lack of ratings below three, with some users barely giving any movies a score of one or two. This is potentially due to most users saving exceedingly negative feedback for the truly disappointing movies. Even though very reasonable realistically, it can introduce unwanted bias into our model, since the value we aim to predict is actually the rating. For example, because it has seen too little examples of low ratings, our model might be inclined to ignore these, and thus is less able to capture the full nuance of rating prediction. To remedy this, we duplicated all training data points with a rating value below three. This step is done after computing ANI values, so it does not interfere with any other calculations. Figure 6 also illustrates another characteristic of consumer rating behavior: that different consumers tend to rate differently. User one gave significantly fewer low scores than the other two shown (no one's or two's, and very few three's, which is the most common rating for the other two). This indicates that user one is much more lenient, but also shows how inputting ratings on this simple scale is problematic. It does not take into account how harsh users are and treats all scores on the same scale. For example, a rating of five from a user who regularly gives one's or two's mean much more for prediction than a five from a user who only gives five's. Even though user demographics capture this to some degree, we cannot expect two socially similar users to also be equally harsh with ratings. Therefore, for each rating, instead of passing it in on the 1-5 scale, we pass in how this score differs from the mean rating this user gives according to this equation:

$$n(r) = r - R_{mean}$$

where $r \in R$ denotes any given rating, and $R_{mean}$ denotes the mean rating of the user.

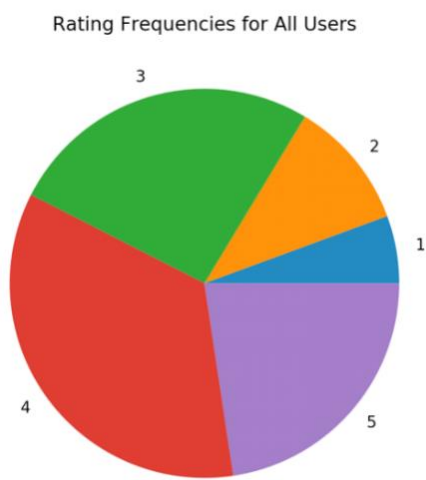

Figure 5: Frequency of user ratings for all users in the ml1-m dataset

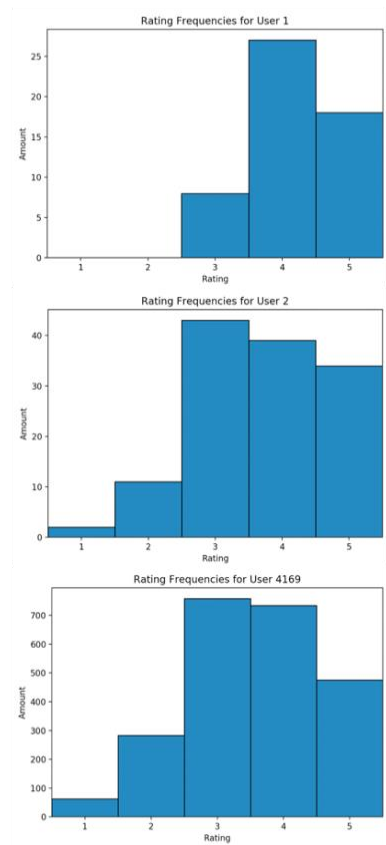

Figure 6: Rating frequencies for users one, two and 4169 (who has the longest action sequence)

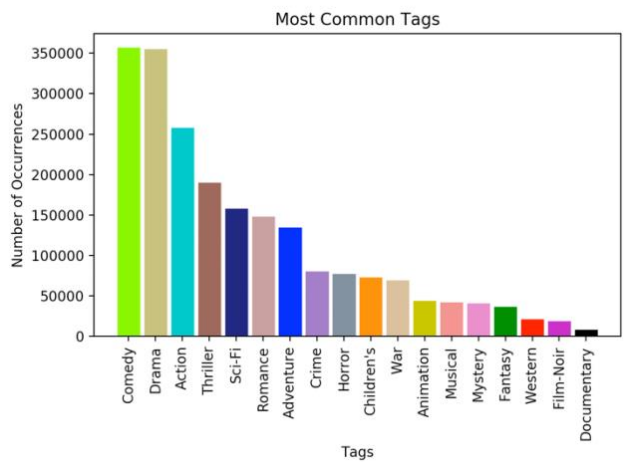

Figure 7: Most common tags for movies in the ml-1m dataset

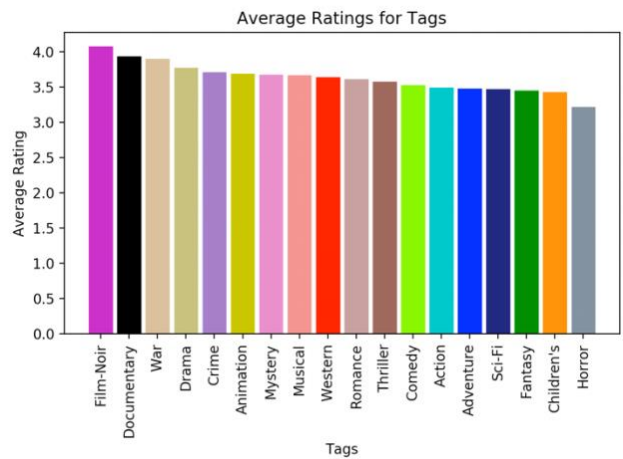

Figure 8: Average ratings for movies containing a given tag

The movie subset also contains interesting information. Figure 7 shows that this particular dataset has a large number of movies tagged drama and comedy, and that the tag frequencies form a long tail distribution when sorted. These two genres have far more entries than any other genre. However, as shown in Figure 8, the average ratings each tag got is relatively similar, and there is no strong correlation between tag frequency and average tag rating. This implies that that high-frequency tags are evenly distributed among the movies. Figure 8's ratings are computed as an unweighted average from the ratings of all movies containing the tag. The reasoning behind such an approach as opposed to a weighted average is that some of these genres overlap, and a movie containing a single tag does not necessarily mean that it belongs solely to that one genre.

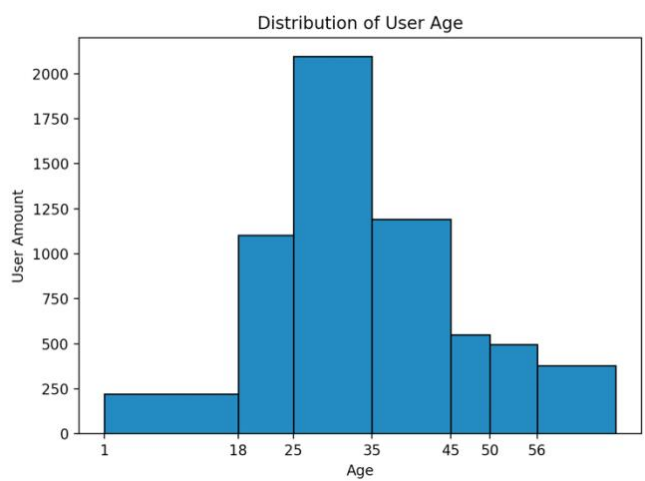

Figure 9: Distribution of user ages

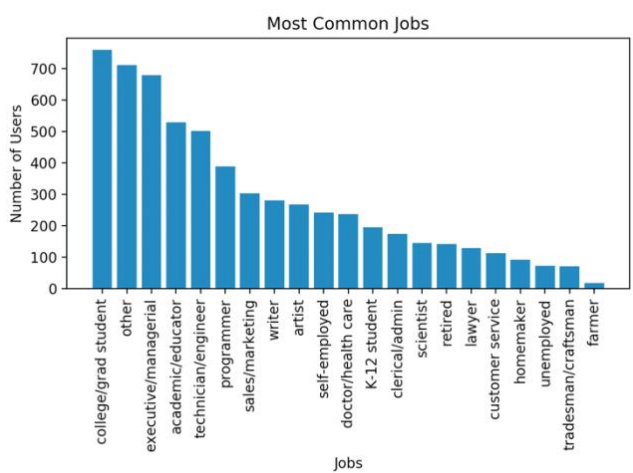

Figure 10: Most common jobs among users

Lastly, we also analyzed the data subset on users. Table 2 shows the male to female ratio among users. There are significantly more men in the ml-1m dataset than women. Even though this might add bias, duplicating rows where the user is female didn't increase the metrics score by much. The lack of increase might be caused by how there still is an incredibly large amount of ratings given by women, and the model is already very capable of generalizing from these. Figure 9 shows the distribution of user ages, and from it we can see that the most active movie-viewing users are aged 25-35, and that overall it is a slightly skewed distribution. Figure 10 illustrates the most common jobs of all the users. Surprisingly, college/grad students comprise the largest group of movie-viewers on the ml-1m dataset. This might be caused by how most users aged 18-25 are college students exclusively, while users of higher ages might have a number of different jobs. Reassuringly, the shapes of the distributions of user age and jobs are incredibly similar for all users and all ratings given (graphs of the latter are not shown because they mostly overlap). This indicates that there is no specific group of users giving most of the ratings, and that ratings given by each demographics group is well correlated with the size of that group.

## 4.2 Model Evaluation

In this subsection, we first define the metric used to score the final predictions (nDCG). We then compare DFNSM to NSM and a random predictor according to this metric.

### 4.2.1 The nDCG Metric

Many next-action recommendation models output a list of top-p probable next actions selected from all possible item choices, ranked in order of likelihood of being chosen. In order to score these outputs, several metrics have been defined measuring how high up a model ranks the actually chosen item, if it even is in the top-p list. Among them is the Normalized Discounted Cumulative Gain (nDCG@p) metric, which is widely used and accepted in the field of recommendation. In our case, since the amount of total actions is incredibly large, and the user only makes one choice at a time, we have opted to run the metrics on the entire ranked list of actions (denoted nDCG@all) for both our model and the baseline models, not only the top-p. nDCG@all is given as

$$nDCG@all = \frac{DCG_{all}}{IDCG_{all}}$$

where

$$IDCG_{all} = 1 \text{ and } DCG_{all} = \sum_{i=1}^{all\ actions} \frac{2^{rel_i} - 1}{log_2(i+1)}$$

where $rel_i$ is a binary value that is true when the $i^{th}$ item is the item chosen by the user and false otherwise. The ideal $DCG_{all}$ ($IDCG_{all}$) is one because at any time, the user only takes one action and $rel_i$ will only be true once the entire summation.

### 4.2.2 Comparison of Performance

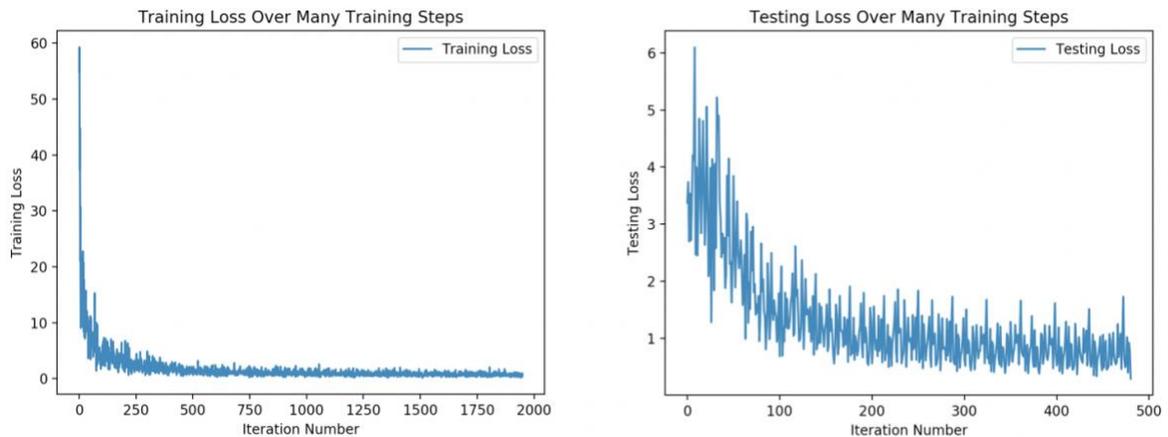

Figure 11: Training and testing loss for the model

For performance evaluation, we use the first twenty users of the ml-1m dataset, and compute nDCG@all for their respective final actions taken. We used more than one user because different users have very different habits and behaviors, so using many people to evaluate the models is more accurate. These values are shown in Figure 12, and their means are shown in Table 3. We compare our model against the following baselines:

- Novelty-Seeking Model (NSM): A model which also leverages novelty-seeking information to predict the likely next action [2].

- Random Predictor: A model that randomly ranks the possible actions for its prediction. (Note shown in Figure 12 because values don't vary much)

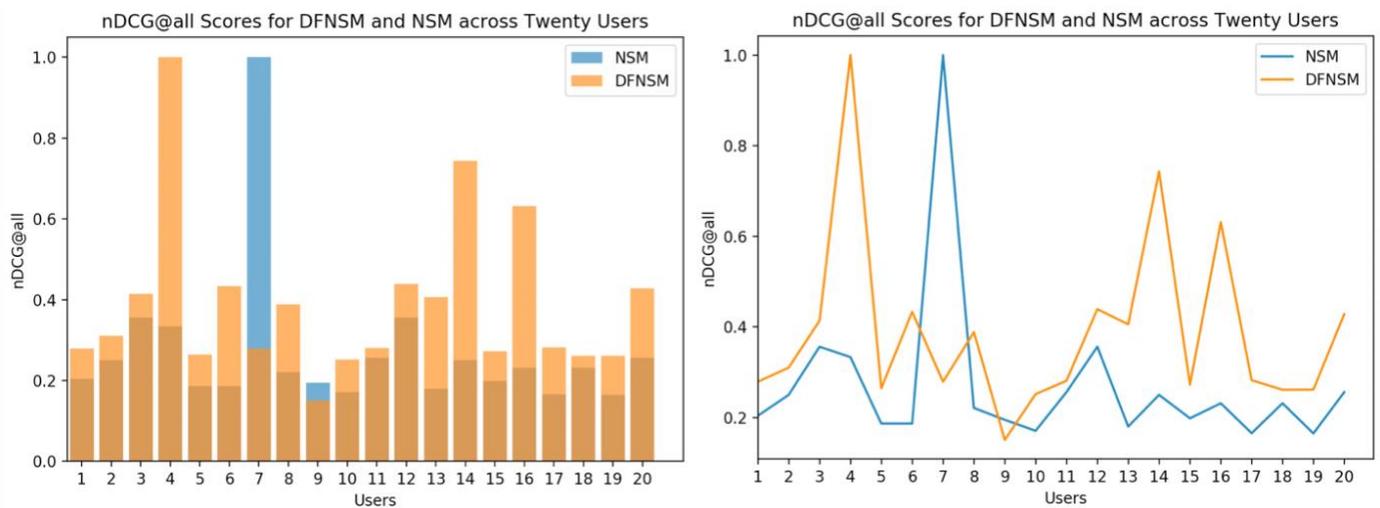

Figure 12: Two representations of the nDCG@all metric for DFNSM and NSM

| Table 3: Mean nDCG@all Metric Scores for DFNSM and NSM | | |
|---|---|---|
| DFNSM | NSM | Random Predictor |
| 0.3887 | 0.2696 | 0.1781 |

From Figure 12 and Table 3, we can see that both DFNSM and NSM outperforms the random predictor. DFNSM also scores better than NSM on average and for the majority of users. NSM only surpasses DFNSM on users seven and nine. These results show us that the many improvements made in DFNSM (defining ANI, only observing past k actions, deep learning integration, etc.) drastically improved its performance.

### 4.3 Observations about UNI

When formulating novelty-seeking, only the action novelties are used for prediction, because user novelties at different times is entirely derived from ANI values. UNI values are therefore only for interpretation purposes. Figure 13 shows the UNI values for five different users at different steps in their action sequences with a k-value of twenty. Note that some lines stop before reaching the edge of the graph window because their user's action sequences are shorter. From these, we can observe several interesting trends.

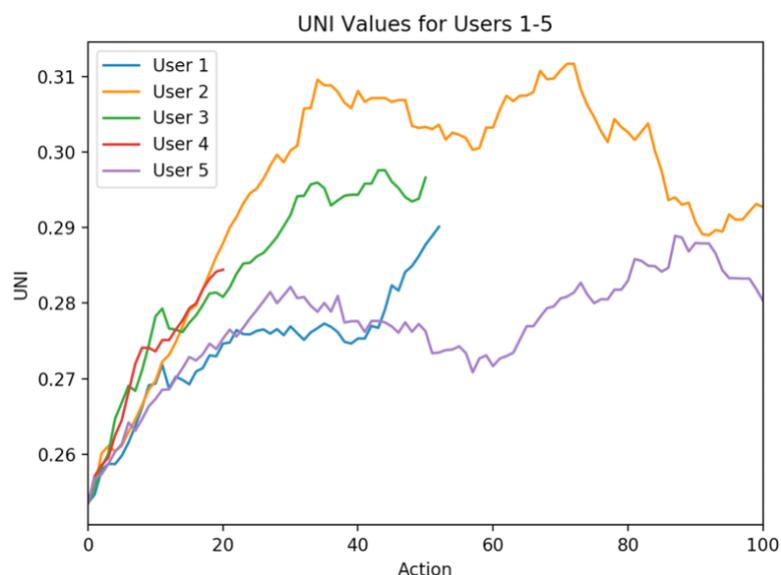

Figure 13: UNI values for five users over 100 actions when k=20

Firstly, the UNI values all start by increasing, and the points where they start to stabilize depends heavily on the user. The first part is reasonable because at the start of every action sequence all no choices have been made, and thus action novelties at that time is going to be completely even, which produces the maximum entropy. The reason why some UNI values tend to stabilize after others might lie in the fact that some users like to explore more varied tags before find a few they are interested in. Secondly, some users are significantly more novelty-seeking than others. For example, user two's UNI consistently dwells above that of user five. Note here that each user's long-gone actions have no impact on their current ones, which means that user two is consistently staying novelty-seeking. This phenomenon means that user novelty-seeking is in fact a factor that is very distinguishable between users, and further reinforces its importance when used for recommendation.

## 4.4 Optimal k-value

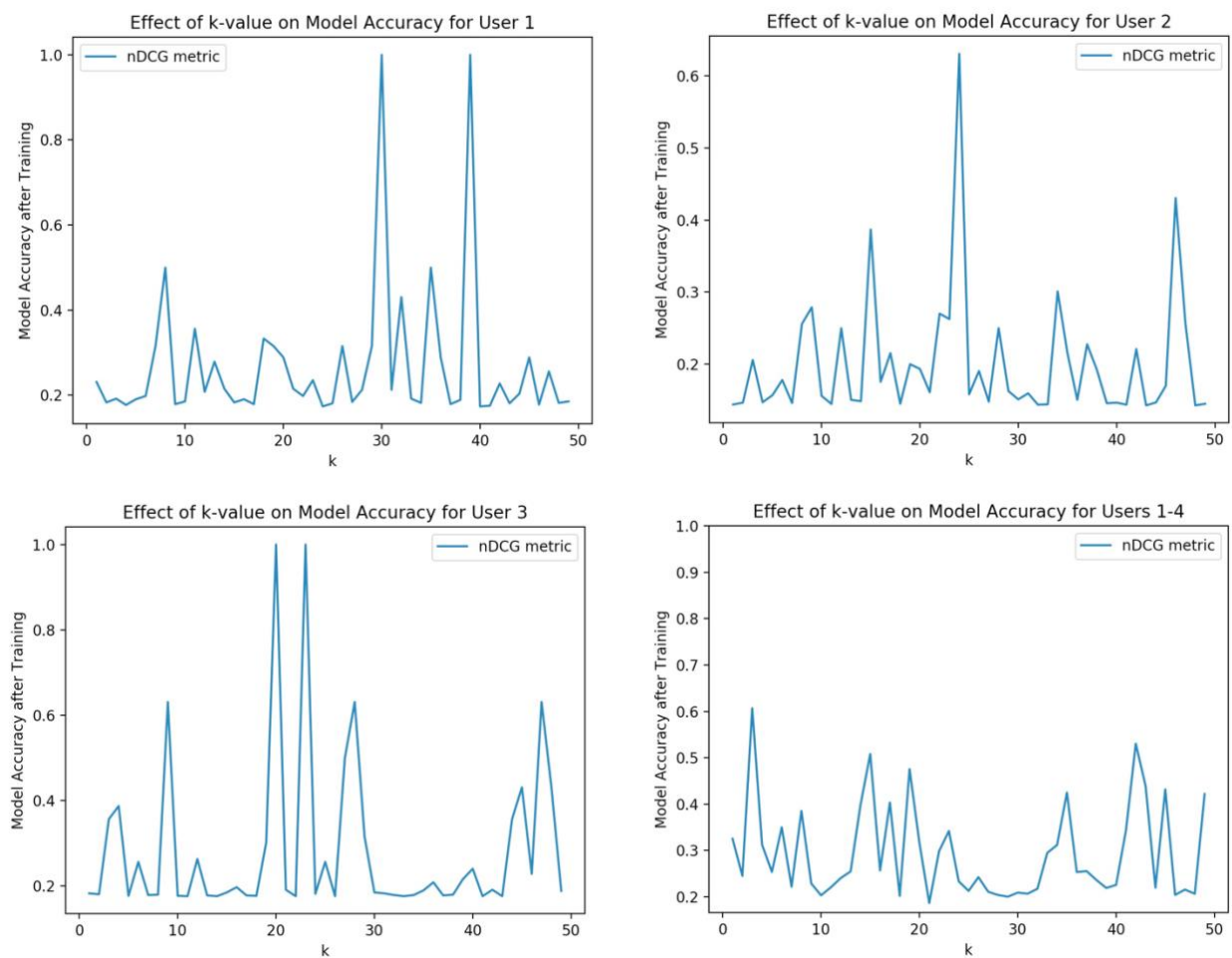

Figure 14: The effects of k on model performance

As explained in the previous section, the amount of previous actions used to compute ANI values is defined as k. This k value can also be thought of as how good we assume a user's memory of past actions is. The larger the value, the further our model looks back to determine if a given tag will be novel or not, and the better we assume the user's memory to be. From this, the value of k that produces the best overall metrics score is also the best approximation of this particular person's memory. Note that this use of the word "memory" differs from traditional interpretations, as we take it mainly to mean how slowly past actions stop affecting current ones.

In this experiment, we trained the Deep Forgetful Novelty-Seeking Model many times for the first three users of the ml-1m dataset separately, and then we tried defining one k value for the first four users. Each training session we select a different k-value and record the overall score the model got when ran through the nDCG metric. The results are shown in Figure 14. These results could have been slightly influenced by more factors than just k, because the initializers for model parameters are drawn randomly. However, when ran multiple times, the overall trends are consistent.

When done separately for the three users, our model produced noticeable spikes in accuracy, and most of the time these spikes are clustered around a small range of k-values. However, when the model is trained with one k-value on multiple users (the graphed nDCG@all value is averaged over all users tested), it was never able to achieve a high metric score. These would indicate that an optimal k-value exists but is different over many users and shows that different users have different memories. Also notice that the spikes in accuracy do not always occur when k is a very large value, thereby confirming the initial assumption that not all previous actions are needed for prediction, and that most people do not have a perfect memory of past actions.

# 5 Conclusion

This paper proposes a next-action recommender system known as Deep Forgetful Novelty-Seeking Model (DFNSM), which predicts on the ml-1m dataset the most likely movies a user might watch next. It is based on the assumption that users do not have perfect memories of past actions, and that long-ago actions do not strongly impact current ones. We first designed a new way to measure novelty for both actions and users, and then we used deep learning to make the final predictions. Extensive experiments conducted on DFNSM showed the validity of our assumptions and also demonstrated the effectiveness of our model compared to other ones. Further research might aim to better determine or predict forgetfulness, which we termed k. Ultimately, our research sheds light on how the phycological phenomenon of forgetfulness impacts user actions, and how it should not be ignored when modeling novelty-seeking.

# 6 Acknowledgements

I would like to express my gratitude to my mentor, Yingmin Zhou, for giving me incredibly helpful advice in times of difficulty and confusion, as well as guiding my initial explorations in the field of recommendation. I would also like to thank Baoyi Chen, who inspired my interest in recommender systems and gave me his support throughout the entire research.